\documentclass[aip,jap,reprint,groupedaddress]{revtex4-1}

\usepackage{graphicx}
\usepackage{upgreek}

\begin{document}

\title{Weak localization and Raman study of anisotropically etched graphene antidots}

\author{Florian Oberhuber}
\author{Stefan Blien}
\author{Stefanie Heydrich}
\author{Fatemeh Yaghobian}
\author{Tobias Korn}
\author{Christian Sch\"uller}
\author{Christoph Strunk}
\author{Dieter Weiss}
\author{Jonathan Eroms}
\email{jonathan.eroms@physik.uni-regensburg.de}

\affiliation{Institut f\"ur Experimentelle und Angewandte Physik, Universit\"at Regensburg, Universit\"atsstr. 31, D-93053 Regensburg, Germany}

\date{\today}

\begin{abstract}
We study a crystallographic etching process of graphene nanostructures where zigzag edges can be prepared selectively. The process involves heating exfoliated single-layer graphene samples with a predefined pattern of antidot arrays in an argon atmosphere at 820\,$^{\circ}$C, which selectively removes carbon atoms located on armchair sites.  Atomic force microscopy and scanning electron microscopy cannot resolve the structure on the atomic scale. However, weak localization and Raman measurements –- which both probe intervalley scattering at armchair edges –- indicate that zigzag regions are enhanced compared to samples prepared with oxygen based reactive ion etching only.
\end{abstract}



\maketitle

Graphene nanoribbons (GNRs) \cite{nakada1996,son2006prl,son2006nature}, as well as step edges on highly ordered pyrolytic graphite (HOPG) \cite{kobayashi1993,klein1994,fujita1996}, were predicted to show a high local electronic density of states, if these edges are along the crystallographic zigzag orientation. For HOPG this was demonstrated experimentally by scanning tunneling microscopy and spectroscopy \cite{kobayashi2005,niimi2005}.
An exciting property of the zigzag edge, which still remains to be confirmed experimentally, is the spin-polarized edge state \cite{son2006prl,son2006nature,wimmer2008}. Its observation is a challenging task, since edges cannot be defined by electron beam lithography (EBL) \cite{han2007} with atomic precision. Typical bottom-up processes, which, for instance, rely on organic molecules \cite{cai2010} or growth on templated SiC surfaces \cite{sprinkle2010}, and those which are based on breaking up carbon nanotubes (CNTs) \cite{jiao2009} can create crystallographically defined edges. These approaches, however, do not allow position control to an extent comparable to lithographic methods.

Crystallographically anisotropic etching is selective concerning the etching of carbon atoms located at different edges, and bears the potential to define armchair or zigzag edges without roughness on the atomic scale \cite{datta2008,ci2008,campos2009,gao2009,nemesincze2010,krauss2010,xie2010,yang2010,shi2011,diankov2013}. The etching reaction can occur in a catalyzed \cite{datta2008,ci2008,campos2009,gao2009}, or in a non-catalyzed form \cite{nemesincze2010,xie2010,yang2010,shi2011,diankov2013}. Only the latter allows position control of the edges via a lithographic patterning process prior to the anisotropic etching \cite{nemesincze2010,xie2010,yang2010,shi2011}. Quality demonstration of edges obtained by anisotropic etching has focused on Raman spectroscopy \cite{nemesincze2010,krauss2010,xie2010,yang2010,shi2011} and electron transport measurements of the ambipolar field effect \cite{xie2010,yang2010,shi2011}. The interpretation of Raman spectra acquired on crystallographically defined edges is based on the edge-orientation-sensitive elastic intervalley scattering of charge carriers between the K and K' valleys. Another way to probe intervalley scattering are electron transport measurements of the weak localization (WL) feature \cite{mccann2006}, however, to this day, no experimental data has been reported on anisotropically etched graphene.

We performed a crystallographically anisotropic carbothermal etching process similar to that reported in [\onlinecite{nemesincze2010}] on graphene antidot lattices \cite{eroms2009njp,heydrich2010} patterned by EBL and reactive ion etching (RIE) with an oxygen plasma. After discussing our carbothermal anisotropic etching process, we will present a comparative study of Raman and WL measurements on a set of samples with focus on intervalley scattering. Part of the samples in this study was not subjected to the anisotropic etching process, but only patterned by EBL and RIE \cite{eroms2009njp,heydrich2010}. By analyzing the intervalley scattering process in both types of samples by both methods, we can deduce that our anisotropic etch step predominantly generates zigzag edges, however, with varying edge roughness, which is still not well controlled.

\begin{figure}
\includegraphics{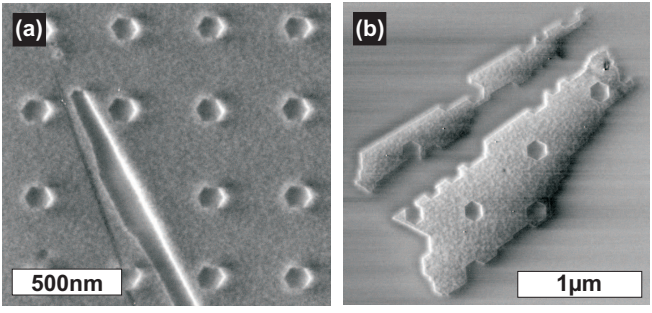}
\caption{\label{fig_semimages}SEM images of single- (a), and bilayer graphene (b) after application of the preconditioning step and conducting the etching reaction. Anisotropic etching generated the hexagonal shape of the antidots and the flakes' straight borders.}
\end{figure}

Samples were prepared by exfoliation of natural graphite onto Si chips covered with a 300\,nm oxide layer. By EBL and RIE, we defined square lattices of circular holes with diameter $d$\,$\sim$\,40\,nm and lattice constants $a $\,$ \sim$\,140\,-\,450\,nm in the flakes. In a manner similar to [\onlinecite{nemesincze2010}], samples were etched at a temperature $T$\,$ \sim$\,820\,$^{\circ}$C in a quartz tube reactor in a flow of Ar gas (purity$\, \geq \,$99.9999\,\%, O$ _2 $\,$\leq$\,0.5\,ppm) at ambient pressure. During this anisotropic etching step the antidots were grown from a circular to a hexagonal geometry and to diameters $d$\,$\sim$\,100\,-\,150\,nm (cf. Fig\,\ref{fig_semimages}). The samples on which we did not perform anisotropic etching were prepared with the same lattice constants, but antidot diameters $d $\,$ \sim$\,40\,-\,165\,nm. We performed scanning electron microscopy (SEM) and atomic force microscopy (AFM) to characterize the etching process. In order to keep the samples clean, care was taken not to expose the flakes to the electron beam of the SEM before conducting Raman or transport measurements. Raman spectra were recorded with 532\,nm circularly polarized light. Details of the Raman setup are reported elsewhere \cite{yaghobian2012}. In order to perform four-point transport measurements, flakes were patterned into Hall bars by EBL and RIE, and contacts were fabricated by EBL and evaporation of Pd, Re, or Au with an adhesion layer of Cr or Ti. Transport data were recorded in He-cryostats at an AC current of 10\,nA, while the charge carrier density was controlled via the back gate.

Anisotropic etching showed the best performance in a narrow temperature range around $T$\,$ \sim$\,820\,$^{\circ}$C, with lateral etch rates $\sim$\,20\,$\frac{\mathrm{nm}}{\mathrm{h}}$ for single-layer graphene, where $T$ denotes the temperature of the outer wall of the quartz tube.
In [\onlinecite{nemesincze2010}], the reaction was suggested to occur between graphene and the SiO$_2$'s oxygen atoms. However, several observations in our experiments are incompatible with this scenario. Experiments with multi-layer graphene, and single-layer graphene in UHV or in a H$_2$ atmosphere, suggest that the reaction involves gaseous O$_2$ with concentrations $ \leq \,$0.5\,ppm in the Ar atmosphere\cite{Supplementary}.

For single-layer graphene, we observed a remarkable behavior, in line with reports on anisotropic etching with hydrogen plasma \cite{shi2011,diankov2013}: while etching was observed to be anisotropic for graphene two or more layers thick, it was isotropic for single-layer graphene, unless we applied a specific sample preconditioning step prior to etching. This preconditioning involves heating the samples to $T$\,$ \sim$\,850\,$^{\circ}$C, with a stainless steel grid located upstream of the sample\cite{Supplementary}.

\begin{figure}
\includegraphics[scale=1.0]{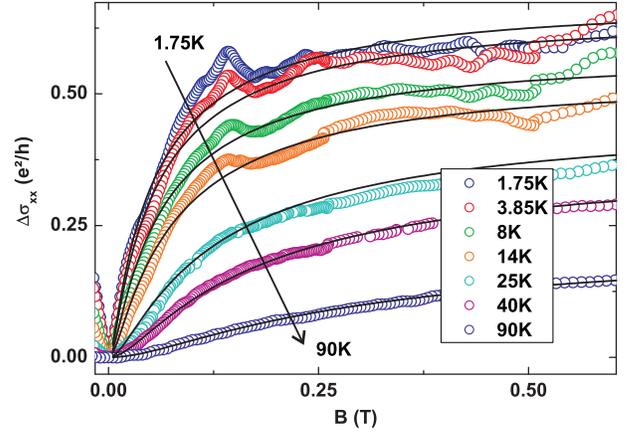}
\caption{\label{fig_wlmeas}Experimental data for different temperatures on the WL peak (circles), measured on a sample with anisotropically etched hexagonal antidots of diameter $d=100$\,nm patterned in a square lattice of constant $a=200$\,nm. The lines display fit curves according to [\onlinecite{mccann2006}].}
\end{figure}

Now, we discuss the quality of anisotropically etched samples by analyzing the WL. WL is a phase-coherent effect whose sign and amplitude in graphene depends on the interplay of the phase-coherence length, the intervalley scattering length and the intravalley scattering length. In particular, intervalley scattering occurs at armchair edges, making the effect visible\cite{morozov2006,saito2001,cancado2004}. We measured WL for samples that had undergone anisotropic etching, and samples that had not, i.e., we compare WL for samples with lattices of hexagonal and circular antidots, respectively. In order to extract intervalley scattering lengths from the WL peaks, we fitted the raw data to the theory \cite{mccann2006} according to the equation
$
\delta \sigma (B) = \frac{e^2}{\pi h} \left[  F\left( \frac{B}{B_\varphi} \right)  -  F\left( \frac{B}{B_\varphi + 2B_i} \right)  -  2F\left( \frac{B}{B_\varphi + B_\ast} \right)  \right]
$ for single-layer graphene, as shown in Fig.\,\ref{fig_wlmeas}. Here, $
F (z) = \mathrm{ln}(z) + \psi \left( 0.5 + z^{-1} \right)
$, $
B_{\varphi , i , \ast} = \frac{\hbar}{4De}\tau_{\varphi , i , \ast}^{-1}
$, and $
\tau_\ast^{-1} = \tau_w^{-1} + \tau_z^{-1} + \tau_i^{-1}
$. $\psi$ is the digamma function, $D$ the diffusion constant, $\tau_\varphi$ denotes the phase coherence time, $\tau_i$ the intervalley scattering time, and $
\left(\tau_w^{-1} + \tau_z^{-1}\right)^{-1}$ the intravalley scattering time. The scattering lengths are related to the scattering times via $L_{\varphi , i , \ast} = \sqrt{  D \, \tau_{\varphi , i , \ast}  }$. Performing the measurements at different temperatures allowed us to prove WL as the origin of the observed peaks, and to extract a $L_\varphi \propto T^{-0.5}$ dependence \cite{eroms2009njp}.

\begin{figure}
\includegraphics{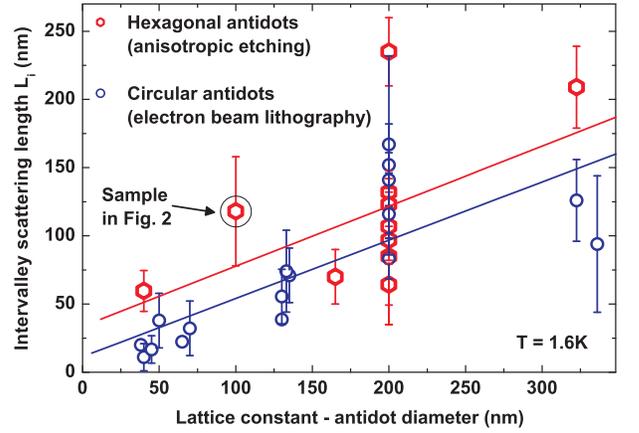}
\caption{\label{fig_wldata}$L_i$ for a set of single-layer graphene samples with circular (blue color) and a set with anisotropically etched hexagonal antidots (red). Part of the samples with circular antidots was already reported in [\onlinecite{eroms2009njp}]. Lines in respective color display fits to each data set. The data point highlighted by a black circle around it stems from the measurement shown in Fig.\,\ref{fig_wlmeas}.}
\end{figure}

\begin{figure}
\includegraphics{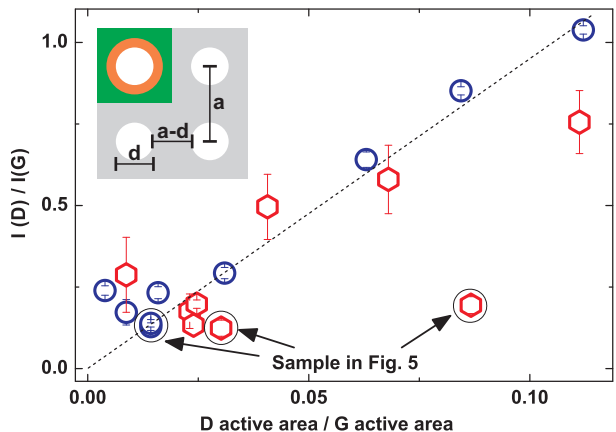}
\caption{\label{fig_ramandata}Integrated intensity ratios $I(D)/I(G)$ vs. the ratio of peak activation areas. The dashed line is a guide to the eye. Legend and origin of the data are as in Fig.\,\ref{fig_wldata}. The inset shows a schematic of graphene (grey) patterned with an antidot lattice (white) with constant $a$ and antidot diameter $d$. The orange-colored area displays the activation area for the D peak, the orange and green-colored areas together show the activation area of the G peak within one unit cell of the lattice, respectively. The data points highlighted by black circles stem from the spectra shown in Fig.\,\ref{fig_ramanevol}.}
\end{figure}

Circular antidots would be expected to have roughly an equal amount of armchair and zigzag segments. Hexagonal antidots created by anisotropic etching can be terminated by zigzag or armchair edges. For antidots with zigzag edges, a lower amount of intervalley scattering events leads to an increased $L_i$, and vice versa for armchair edges. In Fig.\,\ref{fig_wldata} we plot $L_i$ extracted from data at $T$\,$=$\,1.6\,K for different single-layer graphene flakes with hexagonal, or circular antidots \footnote{Strictly speaking, WL theory \cite{mccann2006} was derived for homogeneously distributed defects in graphene}. Different data points in part stem from different samples, in part from samples which were measured at different charge carrier densities. In [\onlinecite{eroms2009njp}] we found a linear relationship between $L_i$ and the spacing $a-d$ between neighboring antidots (see sketch of the lattice in inset of Fig.\,\ref{fig_ramandata}), which can also be seen in Fig.\,\ref{fig_wldata} for the data on circular antidots (blue data points). The data for anisotropically etched graphene shows significant scatter. The highest values for $L_i$ at given values for $a-d$ are observed for anisotropically etched samples, and reach up to 235\,nm, whereas the lowest values for $L_i$ are predominantly observed for samples with circular antidots. For comparison, values for $L_i$ reported for graphene on SiO$_2$ without intentional defects lie in the range 250\,nm - 1\,$\upmu$m \cite{gorbachev2007,tikhonenko2008prl}. From this, we deduce that our anisotropic etching reaction favors the creation of zigzag edges, however, with varying edge quality. With our available microscopy methods, we could not resolve differences between anisotropically etched samples with comparatively high or low values for $L_i$. Hence, the anisotropic etching process must generate an edge roughness on a scale lower than our microscopy resolution limit of $\sim$\,3\,nm for AFM and $\sim$\,1.5\,nm for SEM.

In addition to the WL data, we studied Raman spectra on single-layer graphene with anisotropically etched hexagonal antidots, and single-layer graphene with circular holes. Intervalley scattering is responsible for the appearance of the defect, or D peak, in the spectrum of sp$^2$-hybridized carbon \cite{saito2001,cancado2004}. If there are no other significant scattering sources, such as defects, on the sample, the D peak intensity measures the amount of armchair edges within the area of the graphene flake, that is illuminated by the laser beam \cite{maultzsch2004}. Anisotropically etched graphene edges have been studied with regard to the ratio of the D peak intensity $I(D)$ over the G peak intensity $I(G)$ \cite{nemesincze2010,krauss2010,xie2010,yang2010,shi2011}. The G peak intensity measures the amount of sp$^2$-hybridized carbon atoms within the illuminated sample area. Perfect zigzag edges cannot generate a D peak, and, therefore, the expected ratio $I(D)/I(G)$ would be zero. The ratio $I(D)/I(G)$ for circular holes with a roughly equal amount of zigzag and armchair segments would be expected to lie between that for zigzag and armchair edges. Further, the D and G peak intensities not only depend on the microscopical composition of the edges, but also on the polarization of linearly polarized light with respect to the edge orientation \cite{cancado2004july,cancado2004,gupta2009}. The use of circularly polarized light in our setup, however, eliminates this effect.

\begin{figure}
\includegraphics{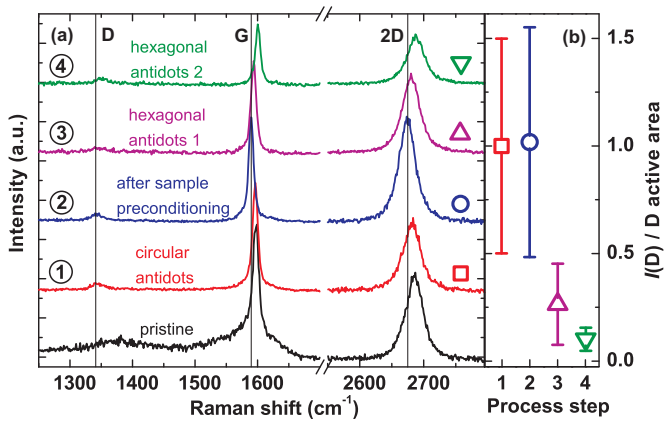}
\caption{\label{fig_ramanevol} (a) Raman spectra of a single-layer graphene flake between different stages of sample processing. Antidots were patterned in a square lattice of constant $a$\,$=$\,200\,nm. The diameter $d$ for round antidots was $\sim$\,40\,nm, preconditioning left it unchanged, and they were grown to $\sim$\,80\,nm and $\sim$\,150\,nm within the first and second anisotropic etching step, respectively. The three vertical lines mark the positions for the D, G and 2D peak for the spectrum after the sample preconditioning step. The symbols on the right-hand side refer to the data points in Fig.\,\ref{fig_ramanevol}\,(b). (b) Plot of the normalized ratio between D peak intensity and the D active area for the spectra in (a) in their respective symbol and color.}
\end{figure}

In order to compare samples with different antidot lattice geometries, we consider their unit cell. Several antidot lattice unit cells are contained within the laser spot diameter of about 1\,$\upmu$m. In each of them, the G peak is generated in the area covered by carbon atoms, i.e., $a^2 - (d/2)^2\pi$. The mechanism for D peak Raman scattering involves a virtual state, which gives the process a certain lifetime related to a length scale of $\sim\,$4\,nm via the Fermi velocity \cite{casiraghi2009nl,lucchese2010,jorio2010}. Consequently, the D peak in each unit cell should be generated in an area formed by a ring of width $\sim$\,4\,nm around the antidot, i.e. $[(d/2+4\,\mathrm{nm})^2 - (d/2)^2]\pi \; \sim \; d \pi \, \times \, 4\,\mathrm{nm}$ (cf. Fig.\,\ref{fig_ramandata}).
The ratio $I(D)/I(G)$ should be proportional to the ratio of the areas activating the respective peaks. In Fig.\,\ref{fig_ramandata} we plot the integrated intensity ratio \footnote{In the literature, peak intensities account for the peak heights, or the integrated Lorentz areas underneath the peaks. Except for the high defect density regime, both methods give the same $I(D)/I(G)$ intensity ratios \cite{jorio2010}.} vs. the peak activation area ratio for different samples. As expected, for samples with round holes the graph shows a linear relationship (blue data points). Compared to circular antidots, $I(D)/I(G)$-values for anisotropically etched antidots lie in the same range, or lower, at respective peak activation area ratios (red data points). As for WL, the edge quality of samples with comparatively high or low $I(D)/I(G)$-values cannot be distinguished with our microscopic methods. This supports the conclusions that we drew from the WL data. The spectra for the anisotropically etched samples in Fig.\,\ref{fig_ramandata} with the lowest ratio of $I(D)/I(G)$ were acquired on one and the same flake and are displayed in Fig.\,\ref{fig_ramanevol}.

The Raman spectra shown in Fig.\,\ref{fig_ramanevol}\,(a) were recorded on a single-layer graphene flake between consecutive sample processing steps. Spectra were acquired for the sample in its pristine state, after definition of circular antidots by EBL and RIE, after sample preconditioning, and after two separate etching steps. The corresponding $I(D)/I(G)$-values in Fig.\,\ref{fig_ramandata} are marked by black circles and indicate high edge quality after anisotropic etching. Since the lattice constant remains unchanged after the definition of antidots, not only the $I(D)/I(G)$-ratio is a viable measure of the edge quality, but also the ratio of the integrated D peak intensity over the D active area. Fig.\,\ref{fig_ramanevol}\,(b) plots this ratio for four of the spectra in Fig.\,\ref{fig_ramanevol}\,(a). The data in Fig.\,\ref{fig_ramanevol}\,(b) were normalized to the data point for circular antidots. According to the positions for the G and 2D peak\cite{Supplementary}, the pristine flake shows a very high degree of doping \cite{pisana2007,yan2007,casiraghi2009pss,heydrich2010}. Introducing circular antidots decreased the FWHM of the D and G peak, and shifted the D, G, as well as the 2D peak as an indication of a decreased degree of doping. This might be due to a reduction of dopant adsorbates by the liquid solvents involved in the processing. The preconditioning step left the peak intensities unchanged, however, the peak shifts monitor a decrease of doping, which might be explained by the evaporation of dopant adsorbates from the flake at the elevated temperature. The reduction of the G peak intensity after consecutive anisotropic etching steps reflects the reduced carbon-covered area. Further, the first anisotropic etching step induced a reduction of the D peak intensity, as well as a reduction of the ratio of intensity over active area to roughly 25\% compared to the spectrum for the circular antidots. However, application of a second anisotropic etching step to grow the antidots further led to an increase in D peak intensity, which should not occur for perfect zigzag edges. On the other hand, while the absolute D peak intensity increases, the ratio of intensity over active area further decreases due to this second anisotropic etching step, as displayed in Fig.\,\ref{fig_ramanevol}\,(b). The successive decrease of the ratio of D peak intensity over D active area in Fig.\,\ref{fig_ramanevol}\,(b) due to anisotropic etching favors zigzag as dominating edge type. The increase of D peak intensity due to the second anisotropic etching step, as well as the fact that D peak intensities are non-zero, are in line with edge roughness.

\begin{figure}
\includegraphics[width=45mm]{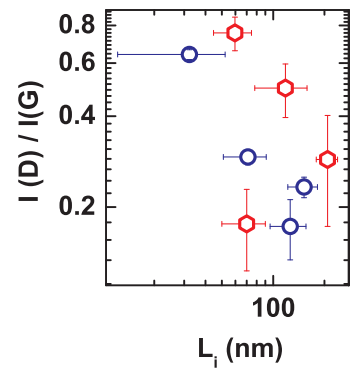}
\caption{\label{fig_ramanwl_correlation}Integrated intensity ratios $I(D)/I(G)$ vs. $L_i$. Round (blue) and hexagonal (red) symbols refer to round and anisotropically etched antidots, respectively. Each data point displays a separate flake, on which both, WL and Raman measurements, were performed.}
\end{figure}

Finally, since both WL and Raman are sensitive to intervalley scattering we plot the Raman D-peak intensity vs. $L_i$ from the WL fit for samples on which both measurements were performed. Fig.\,\ref{fig_ramanwl_correlation} indeed shows a correlation between those quantities\cite{Supplementary}. The large scatter in the data might be explained by considering that the WL-theory in [\onlinecite{mccann2006}] was derived for randomly distributed defects and its applicability is limited for antidot lattices. Another reason could be that Raman spectroscopy probes intervalley scattering at much higher charge carrier excitation energies compared to WL.

In conclusion, we performed anisotropic etching of single-layer graphene on SiO$_2$ at oxygen concentrations of $\sim$\,0.5\,ppm. Anisotropic etching preferentially generated zigzag edges with roughness between the atomic scale and $\sim$\,1.5\,nm. This was demonstrated by studying the intervalley scattering process in WL and Raman measurements for a set of antidot samples, and by tracking the evolution of Raman spectra on a single flake after consecutive processing steps. We also establish a correlation between the D-peak intensity in the Raman spectrum and the intervalley scattering length obtained from fitting the WL data.

\begin{acknowledgments}
We thank Jan Bundesmann for programming a WL fitting tool, Thomas Hofmann for performing the heating experiments in UHV, and gratefully acknowledge financial suppport by the DFG via SFB 689 and GRK 1570.
\end{acknowledgments}

%


\end{document}